\documentclass[sigconf]{acmart}
\usepackage[T1]{fontenc}
\usepackage{lmodern}
\usepackage{algorithm}
\usepackage{algorithmicx}
\usepackage{algpseudocode}
\usepackage{amsmath}
\usepackage{amssymb}
\usepackage{array}
\usepackage{arydshln}
\usepackage{bm}
\usepackage{booktabs}
\usepackage{braket}
\usepackage{enumitem}
\usepackage{graphicx}
\usepackage[mathscr]{euscript}
\usepackage{mathtools}
\usepackage{multirow}
\usepackage{framed} 
\usepackage{nicefrac}
\usepackage{pgfplots}
\usepackage{qtree}
\usepackage{adjustbox}
\usepackage{rotating}
\pgfplotsset{compat = 1.13}
\usepackage{stfloats}
\usepackage{subfig}
\usepackage{url}
\usepackage{tikz}
\usepackage{tkz-berge}
\usetikzlibrary{shapes, backgrounds,calc}
\usetikzlibrary{decorations.pathreplacing}
\usepackage{fancybox}

\tikzstyle{vertex} = [circle, draw, inner sep = 0pt, minimum size = 10pt]
\newcommand{\vertex}{\node[vertex]}

\definecolor{bblue}{rgb}{0.12392, 0.0490, 0.9588}
\definecolor{sskyblue}{rgb}{0.1529, 0.5882, 0.9216}
\definecolor{ggreen}{rgb}{0.5020, 0.7961, 0.3451}
\definecolor{yyellow}{rgb}{0.9765, 0.9804, 0.0784}
\definecolor{rred}{rgb}{1, 0.0, 0.16}
\definecolor{oorange}{rgb}{1.0, 0.55, 0.0}

\setcopyright{acmlicensed}

\begin{document}
\fancyhead{}
\settopmatter{printacmref=false, printfolios=false}

\title{Poster Abstract: Hierarchical Subchannel Allocation for Mode-3 Vehicle-to-Vehicle Sidelink Communications}

\author{Luis F. Abanto-Leon}
\affiliation{%
  \institution{Eindhoven University of Technology}
  \city{Eindhoven} 
  \country{Netherlands} 
}
\email{l.f.abanto@tue.nl}

\author{Arie Koppelaar}
\affiliation{%
	\institution{NXP Semiconductors}
	\city{Eindhoven} 
	\country{Netherlands} 
}
\email{arie.koppelaar@nxp.com}

\author{Sonia Heemstra de Groot}
\affiliation{%
	\institution{Eindhoven University of Technology}
	\city{Eindhoven} 
	\country{Netherlands} 
}
\email{sheemstradegroot@tue.nl}

\renewcommand{\shortauthors}{L. F. Abanto-Leon et al.}

\begin{abstract}
	In this poster we present a graph-based hierarchical subchannel allocation scheme for V2V sidelink communications in Mode-3. Under this scheme, the eNodeB allocates subchannels for in-coverage vehicles. Then, vehicles will broadcast directly without the eNodeB intervening in the process. Therefore, in each communications cluster, it will become crucial to prevent allocation conflicts in time domain since vehicles will not be able to transmit and receive simultaneously. We present a solution where the time-domain requirement can be enforced through vertex aggregation. Additionally, allocation of subchannels is performed sequentially from the most to the least allocation-constrained cluster. We show through simulations that the proposed approach attains near-optimality.
\end{abstract}



\copyrightyear{2017} 
\acmYear{2017} 
\setcopyright{acmlicensed}
\acmConference{SenSys '17}{November 6--8, 2017}{Delft, Netherlands}\acmPrice{15.00}\acmDOI{10.1145/3131672.3136987}
\acmISBN{978-1-4503-5459-2/17/11}

\maketitle
\section{Motivation and Contributions}
In V2V Mode-3, eNodeBs assign subchannels to vehicles in order for them to periodically broadcast CAM messages \cite{b2}. A crucial aspect is to ensure that vehicles in the same cluster will broadcast in orthogonal time subchannels\footnote{A subchannel is a time-frequency resource chunk capable of sufficiently conveying a CAM message.} to avoid conflicts. In general, resource/subchannel allocation problems can be represented as weighted bipartite graphs. However, in this scenario there is an additional time orthogonality constraint which cannot be straightforwardly handled by conventional graph matching methods \cite{b3}. Thus, in our approach the mentioned constraint has been taken into account. We also perform the allocation task in a sequential manner based on the constrainedness of each cluster. To illustrate the gist of the problem, in Fig. 1 we show two partially overlapping clusters where a conflict between vehicles $V_8$ and $V_{10}$ is generated as the allotted subchannels are in the same subframe.
\begin{figure}[t]
	\begin{center}
		\begin{tikzpicture}
		\node (img) {\includegraphics[width=0.96\linewidth]{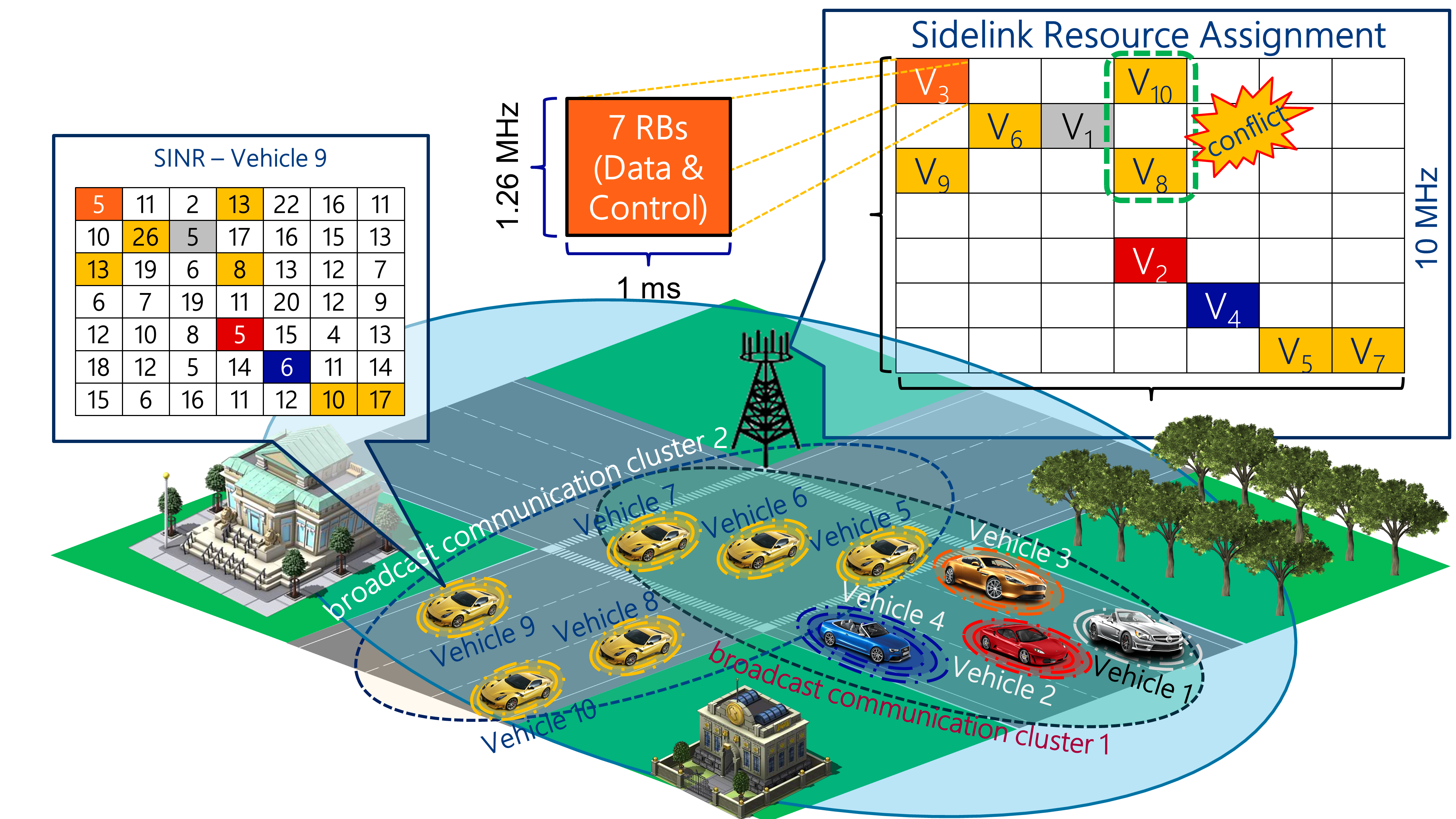}};
		\node[rotate = 90] at (0.67,1) {\tiny $\tiny K$};
		\node at (2.27,-0.07) {\tiny $\tiny L$};
		\end{tikzpicture}
		\caption{V2V broadcast communications scenario}
		\label{Fig1}
	\end{center}
	\label{f1}
	\vspace{-0.2cm}
\end{figure}

\section{Proposed Approach}
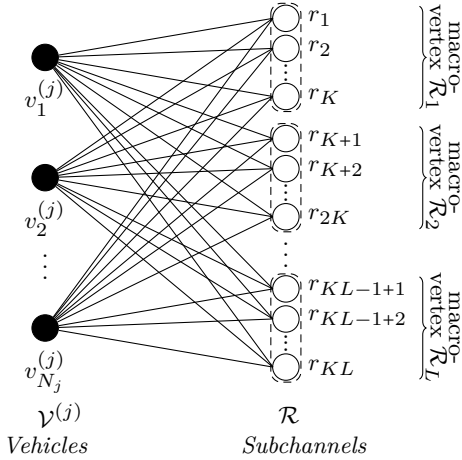
\begin{figure}[t]
	\centering
	\[\begin{tikzpicture}[scale = 0.8]
	
	\vertex[fill] (v1) at (0,-1.25) [label = below:$v^{\text{(}j\text{)}}_{1}$] {};
	\vertex[fill] (v2) at (0,-3.25) [label = below:$v^{\text{(}j\text{)}}_{2}$] {};
	\node at (0,-4.6) {\vdots};
	\vertex[fill] (v3) at (0,-5.75) [label = below:$v^{\text{(}j\text{)}}_{N_j}$] {};
	
	\vertex (r11) at (4,-0.6) [label = right:$r_{1}$] {};
	\vertex (r12) at (4,-1.1) [label = right:$r_{2}$] {};
	\node[rotate = 90] at (4,-1.5) {$...$};
	\vertex (r13) at (4,-1.9) [label = right:$r_{K}$] {};
	
	\vertex (r21) at (4,-2.6) [label = right:$r_{K+1}$] {};
	\vertex (r22) at (4,-3.1) [label = right:$r_{K+2}$] {};
	\node[rotate = 90] at (4,-3.5) {$...$};
	\vertex (r23) at (4,-3.9) [label = right:$r_{2K}$] {};
	
	\node at (4,-4.4) {\vdots};
	
	\vertex (r31) at (4,-5.1) [label = right:$r_{K(L-1)+1}$] {};
	\vertex (r32) at (4,-5.6) [label =right:$r_{K(L-1)+2}$] {};
	\node[rotate = 90] at (4,-6) {$...$};
	\vertex (r33) at (4,-6.4) [label = right:$r_{KL}$] {};
	
	\path
	(v1) edge (3.785,-0.6)
	(v1) edge (3.785,-1.1)
	(v1) edge (3.785,-1.9)
	(v1) edge (3.785,-2.6)
	(v1) edge (3.785,-3.1)
	(v1) edge (3.785,-3.9)
	(v1) edge (3.785,-5.1)
	(v1) edge (3.785,-5.6)
	(v1) edge (3.785,-6.4)
	
	(v2) edge (3.785,-0.6)
	(v2) edge (3.785,-1.1)
	(v2) edge (3.785,-1.9)
	(v2) edge (3.785,-2.6)
	(v2) edge (3.785,-3.1)
	(v2) edge (3.785,-3.9)
	(v2) edge (3.785,-5.1)
	(v2) edge (3.785,-5.6)
	(v2) edge (3.785,-6.4)
	
	(v3) edge (3.785,-0.6)
	(v3) edge (3.785,-1.1)
	(v3) edge (3.785,-1.9)
	(v3) edge (3.785,-2.6)
	(v3) edge (3.785,-3.1)
	(v3) edge (3.785,-3.9)
	(v3) edge (3.785,-5.1)
	(v3) edge (3.785,-5.6)
	(v3) edge (3.785,-6.4);
	
	\draw[densely dashed,rounded corners=4]($(r11)+(-.25,.25)$)rectangle($(r13)+(0.25,-.25)$);
	\draw[densely dashed,rounded corners=4]($(r21)+(-.25,.25)$)rectangle($(r23)+(0.25,-.25)$);
	\draw[densely dashed,rounded corners=4]($(r31)+(-.25,.25)$)rectangle($(r33)+(0.25,-.25)$);
	
	\draw[decoration={brace, raise=5pt},decorate] (6,-0.4) -- node[right=6pt] {} (6,-2.1);
	\draw[decoration={brace, raise=5pt},decorate] (6,-2.4) -- node[right=6pt] {} (6,-4.1);
	\draw[decoration={brace, raise=5pt},decorate] (6,-4.9) -- node[right=6pt] {} (6,-6.6);
	
	\node[rotate=-90] at (6.8,-1.25) {macro-};
	\node[rotate=-90] at (6.8,-3.25) {macro-};
	\node[rotate=-90] at (6.8,-5.75) {macro-};
	\node[rotate=-90] at (6.5,-1.25) {vertex $\mathcal{R}_1$};
	\node[rotate=-90] at (6.5,-3.25) {vertex $\mathcal{R}_2$};
	\node[rotate=-90] at (6.5,-5.75) {vertex $\mathcal{R}_L$};
	
	\node[text width = 0.2cm] at (0,-7.2) {$\mathcal{V}^{\text{(}j\text{)}}$};
	\node[text width = 0.2cm] at (-0.5,-7.7) {\textit{Vehicles}};
	\node[text width = 0.2cm] at (4,-7.2) {$\mathcal{R}$};
	\node[text width = 0.2cm] at (3.4,-7.7) {\textit{Subchannels}};
	
	\end{tikzpicture}\]
	\caption{Constrained weighted bipartite graph}
	\label{f2}
	\vspace{-0.3cm}
\end{figure}

The objective is to find an assignment of subchannels to vehicles such that the system capacity is maximized while respecting the constraints that prevent conflicts. This allocation problem is formulated as
\begin{subequations} \label{e1}
	\begin{gather} 
	\begin{align}
	& {\rm max} ~ {\bf c}^T {\bf x} \\ 
	& {\rm subject~to}~
	{\left[
		\begin{array}{c}
		{\bf I}_{N \times N} \otimes {\bf 1}_{1 \times L} \\
		\hline
		{\bf U}_{J \times N} \otimes {\bf I}_{L \times L} 
		\end{array}
		\right]} \otimes {\bf 1}_{1 \times K} ~ {\bf x} = {\bf 1} 
	\end{align}
	\end{gather}
\end{subequations}
where ${\bf c} \in \mathbb{R}^{M},  {\bf x} \in \mathbb{B}^M$ with $M = NLK$. The amount of clusters is $J$ (each consisting of $N_j$ vehicles, $j=1, \dots, J$); the total number of vehicles is $N$; $L$ is the number of time subframes and $K$ represents the number of subchannels per subframe. ${\bf U}$ is the membership matrix which depicts the association of vehicles to the different clusters. Instead of approaching (\ref{e1}) optimally through exhaustive search, we can solve the allocation in a hierarchical and sequential manner for each cluster $\mathcal{V}^{\text{(}j\text{)}}$ while retaining the solutions from previous allocations. Thus, we solve several subproblems in order of constrainedness, which will lead to a suboptimal solution. Each subproblem can be modeled as a weighted bipartite graph $G^{\text{(}j\text{)}} = \big( \mathcal{V}^{\text{(}j\text{)}}, \mathcal{R}, \mathcal{E}^{\text{(}j\text{)}} \big )$ as shown in  Fig. \ref{f2}, where vehicles and subchannels are modeled as vertices. The subchannels $r_k$ in $\mathcal{R}$ have been assembled into $L$ disjoint groups $\{\mathcal{R}_{l}\}_{l = 1}^L$ called macro-vertices in order to manage the time-domain constraints (vertex aggregation). The edge weight $c^{\text{(}j\text{)}}_{ik} = B \log_2 \big(1 + \mathsf{SINR}^{\text{(}j\text{)}}_{ik} \big)$ represents the achievable capacity of vehicle  $v^{\text{(}j\text{)}}_i \in \mathcal{V}^{\text{(}j\text{)}}$ in subchannel $r_k \in \mathcal{R}$. Thus, the following formulation maximizes the capacity per cluster
\begin{subequations} \label{e2}
	\begin{gather} 
	\begin{align}
	& {\rm max} ~ {\bf c}^T_j {\bf x}_j \\
	& {\rm subject~to}~ 
	{\left[
		\begin{array}{c}
		{\bf I}_{N_j \times N_j} \otimes {\bf 1}_{1 \times L}\\
		\hline
		{\bf 1}_{1 \times N_j} \otimes {\bf I}_{L \times L} 
		\end{array}
		\right]} \otimes {\bf 1}_{1 \times K} ~ {\bf x}_j = {\bf 1}.
	\end{align}
	\end{gather}
\end{subequations}
For completeness, we assume that $N_1 = \dots = N_j = L$, thus \\ $\mathbf{x}_j = \text{[} x^{\text{(}j\text{)}}_{1,1}, \dots, x^{\text{(}j\text{)}}_{L,KL} \text{]}^T$ and $\mathbf{c}_j = \text{[} c^{\text{(}j\text{)}}_{1,1}, \dots, c^{\text{(}j\text{)}}_{L,KL} \text{]}^T$. Also, we claim without a proof due to space limitations, that (\ref{e2}) can be recast as (\ref{e3}) without affecting optimality.
\begin{equation} \label{e3}
\hspace{-1.5cm}
\begin{array}{lclcl}
&& {\rm max} ~ {\bf d}^T_j {\bf y}_j \\
&& {\rm subject~to}~ 
{\left[
	\begin{array}{c}
	{\bf I}_{L \times L} \otimes {\bf 1}_{1 \times L}\\
	\hline
	{\bf 1}_{1 \times L} \otimes {\bf I}_{L \times L} 
	\end{array}
	\right]} {\bf y}_j = {\bf 1}.
\end{array}
\end{equation}
where $ {\bf d}_j = \lim_{\beta \to \infty} \frac{1}{\beta} \overset{\substack{\circ}}{\log} \Big\{({\bf I}_{M \times M}\otimes {\bf 1}_{1 \times K}) \mathrm{e}^{\circ \beta {\bf c}_j} \Big\} $ with $\overset{\substack{\circ}}{\log} \{\cdot\}$ and $\mathrm{e}^{\circ \{ \cdot \} }$ representing the element-wise natural logarithm and Hadamard exponential \cite{b4}, respectively. Note that ${\bf y}_j$ is of lower dimensionality than ${\bf x}_j$ and therefore it can be solved with less complexity.

\section{Simulations}
We consider a 10 MHz channel which is divided into subchannels, each with dimensions of 1ms in time and 1.26 MHz in frequency \cite{b5}. In our model, we assume a CAM message rate of 10 Hz. In Fig. \ref{f3}, we compare 4 different algorithms in base of the average over 1000 simulations. We considered both overlapping and non-overlapping clusters, each with at most $N=100$ vehicles. Through simulations, we show that our scheme can attain near-optimality as its performance is similar to exhaustive search. Considering the \textit{system average rate} criterion, our approach has an advantage over the greedy algorithm. Also, when considering the \textit{worst-rate vehicle}, our proposal excels as it is capable of providing a higher level of fairness. In all cases, the random allocation algorithm is outperformed by the other approaches. 
\begin{figure}[!t]
	\centering
	\begin{tikzpicture}
	\begin{axis}[
	ybar,
	ymin = 0,
	ymax = 9.5,
	width = 8.8cm,
	height = 5.5cm,
	bar width = 10pt,
	tick align = inside,
	x label style={align=center, font=\footnotesize},
	ylabel = {Rate [Mbits / s / subchannel]},
	y label style={at={(-0.06,0.5)}, font=\footnotesize,},
	nodes near coords,
	every node near coord/.append style={color = black, rotate = 90, anchor = west, font = \fontsize{6}{7}\selectfont},
	nodes near coords align = {vertical},
	symbolic x coords = {Highest-Rate Vehicle, System Average Rate, Worst-Rate Vehicle, Rate Standard Deviation},
	x tick label style = {text width = 1.6cm, align = center, font = \footnotesize,},
	xtick = data,
	enlarge y limits = {value = 0.55, upper},
	enlarge x limits = 0.18,
	legend columns=2,
	legend pos = north east,
	legend style={font=\fontsize{6}{5}\selectfont, text width=2.8cm,text height=0.02cm,text depth=.ex, fill = none, }]
	]
	\addplot[fill = bblue] coordinates {(Highest-Rate Vehicle,  8.97) (System Average Rate, 8.24) (Worst-Rate Vehicle, 7.14) (Rate Standard Deviation, 0.477)}; \addlegendentry{Exhaustive Search}
	\addplot[fill = rred] coordinates {(Highest-Rate Vehicle, 8.97) (System Average Rate, 8.11) (Worst-Rate Vehicle, 6.94) (Rate Standard Deviation, 0.553)}; \addlegendentry{Proposed Algorithm}
	\addplot[fill = oorange] coordinates {(Highest-Rate Vehicle, 8.97) (System Average Rate, 7.88) (Worst-Rate Vehicle, 5.35) (Rate Standard Deviation, 0.649)}; \addlegendentry{Greedy Algoritm}
	\addplot[fill = ggreen] coordinates {(Highest-Rate Vehicle, 8.17) (System Average Rate, 4.55) (Worst-Rate Vehicle, 1.43) (Rate Standard Deviation, 1.21)}; \addlegendentry{Random Algorithm}
	
	\end{axis}
	\end{tikzpicture}
	\caption{Vehicles data rate}
	\label{f3}
	\vspace{-0.3cm}
\end{figure}
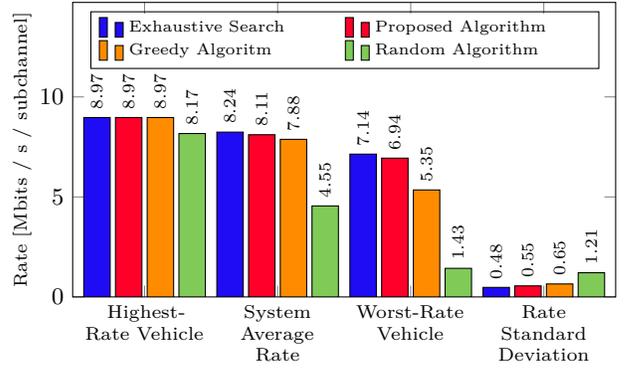
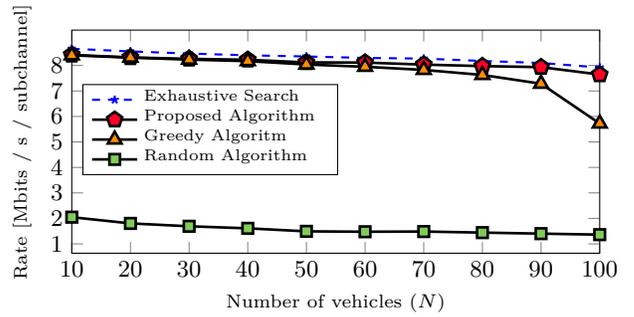
\begin{figure}[!t]
	\centering
	\begin{tikzpicture}
	\begin{axis}[
	xmin = 10,
	xmax = 100,
	width = 8.6cm,
	height = 4.55cm,
	xlabel={Number of vehicles ($N$)},
	xtick={10, 20, 30, 40, 50, 60, 70, 80, 90, 100},
	x label style={align=center, font=\footnotesize},
	ylabel = {Rate [Mbits / s / subchannel]},
	y label style={at={(-0.06,0.5)}, align=center, font=\footnotesize,},
	ytick = {1, 2, 3, 4, 5, 6, 7, 8},
	legend style={at={(0.02,0.35)},anchor=south west, font=\fontsize{6}{5}\selectfont, text width=2.4cm,text height=0.075cm,text depth=.ex, fill = none,},
	]
	
	\addplot[color=blue, mark = star, mark options = {scale = 0.8}, line width = 0.8pt, style = dashed] coordinates 
	{
		(10, 8.6572)
		(20, 8.5512)
		(30, 8.4684)
		(40, 8.3988)
		(50, 8.3525)
		(60, 8.2961)
		(70, 8.2697)
		(80, 8.1770)
		(90, 8.0975)
		(100, 7.9222)
	}; \addlegendentry{Exhaustive Search}
	
	\addplot[color=black, mark = pentagon*, mark options = {scale = 1.5, fill = rred}, line width = 1pt] coordinates 
	{
		(10, 8.4154)           
		(20, 8.3160)     
		(30, 8.2597)    
		(40, 8.2233 )   
		(50, 8.1207)    
		(60, 8.1141)    
		(70, 8.0346)    
		(80, 7.9751)    
		(90, 7.9354)
		(100, 7.6444)    
	}; \addlegendentry{Proposed Algorithm}
	
	\addplot[color=black, mark = triangle*, mark options = {scale = 1.5, fill = oorange}, line width = 1pt] coordinates 
	{
		(10, 8.3988)       
		(20, 8.3094)    
		(30, 8.2299 )     
		(40, 8.1704)     
		(50, 8.0379)       
		(60, 7.9486)   
		(70, 7.8262)   
		(80, 7.6279)      
		(90, 7.2844)   
		(100, 5.7225)     
	}; \addlegendentry{Greedy Algoritm}
	
	\addplot[color=black, mark = square*, mark options = {fill = ggreen}, line width = 1pt] coordinates 
	{
		(10, 2.0484) 
		(20, 1.8033)   
		(30, 1.6912)
		(40, 1.6104)
		(50, 1.4920)
		(60, 1.4766)     
		(70, 1.4845)
		(80, 1.4422)
		(90, 1.4024)    
		(100, 1.3624)
	}; \addlegendentry{Random Algorithm}
	\end{axis}
	\end{tikzpicture}
	\caption{Worst-rate vehicle}
	\label{f4}
	\vspace{-0.3cm}
\end{figure}
Fig. \ref{f4} shows the achievable rate for the \textit{worst-rate vehicle}. We observe that when the vehicle density per cluster is low, the greedy approach attains near optimal solutions as there are more subchannels than vehicles to serve. However, as the density increases, especially near the overload state, its performance dramatically drops. Nevertheless, our approach can provide a fair allocation even in high vehicle density scenarios.

\section{Conclusion}
We have presented a novel hierarchical subchannel allocation scheme for V2V sidelink communications with intra-cluster conflict constraints. The approach is capable of attaining near-optimality at less complexity than exhaustive search.

\bibliography{references} 
\bibliographystyle{unsrt}

%
%
%
%

\end{document}